\documentclass[12pt,a4paper]{article}
\usepackage{fullpage}
\usepackage[centertags]{amsmath}
\allowdisplaybreaks[4]
\usepackage{amsfonts}
\usepackage{amssymb}
\usepackage{bm}
\usepackage{mathrsfs}
\usepackage{url}
\usepackage[dvips,hyperindex]{hyperref}
\begin{document}

\begin{flushright}
\textsf{17 July 2008}
\end{flushright}
\begin{center}
\Large\bfseries
Comment on ``A neutrino's wobble?''
\\[0.5cm]
\large\normalfont
Carlo Giunti
\\[0.5cm]
\normalsize\itshape
\setlength{\tabcolsep}{1pt}
\begin{tabular}{cl}
INFN, Sezione di Torino,
Via P. Giuria 1, I--10125 Torino, Italy
\end{tabular}
\end{center}

\begin{flushleft}
Sir\footnote{
Letter refused by Nature:
\begin{quote}

Dear Dr Giunti

Thank you for your Correspondence submission, which we regret we are unable to publish.
Pressure on our limited space is severe, so we can offer to publish only a very few of the many submissions we receive.

Naturally, I am sorry to convey a negative response in this instance.

Thank you again for writing to us.

Yours sincerely

\ldots

Correspondence
Nature
\end{quote}
},
\end{flushleft}

I think that the ``neutrino's wobble'' explanation of the GSI time anomaly \cite{0801.2079}
reviewed in ``News \& Views'' \cite{Walker-Nature-453-864-2008}
is in contradiction with basic physical principles \cite{0805.0431}.

The observed oscillatory time modulation
of the electron-capture decay rates of
${}^{140}\text{Pr}^{58+}$
and
${}^{142}\text{Pm}^{60+}$
ions
could be quantum beats
due to the existence
of two energy levels of the decaying ions with an
extremely small energy splitting
($ \Delta{E} \approx 6 \times 10^{-16} \, \text{eV} $).
Since the two energy levels develop different phases after the ion creation,
they interfere in the following decay.

On the other hand,
neutrinos are produced in the electron-capture decays
${}^{140}\text{Pr}^{58+} \to {}^{140}\text{Ce}^{58+} + \nu_{e}$
and
${}^{142}\text{Pm}^{60+} \to {}^{142}\text{Nd}^{60+} + \nu_{e}$,
whose rates are measured in the GSI experiment by detecting the ions.
The undetected electron neutrino $\nu_{e}$ is a superposition of massive neutrinos.
In the simplest case of two-neutrino mixing,
we have
$ \nu_{e} = \cos\vartheta \nu_{1} + \sin\vartheta \nu_{2} $,
where $\vartheta$ is the mixing angle and $\nu_{1}$ and $\nu_{2}$
are the massive neutrinos.
Since the two massive neutrinos develop different phases after their creation,
they can interfere after the decay, generating neutrino oscillations.
However,
there cannot be any effect backwards in time on the decay.
Such an effect would violate causality.

Yours sincerely,

\begin{center}
Carlo Giunti
\end{center}

\raggedright

\end{document}